# Quantum Machine Learning for Predicting Binding Free Energies in Structure-Based Virtual Screening

Pei-Kun Yang

E-mail: peikun@isu.edu.tw

ORCID: https://orcid.org/0000-0003-1840-6204

**Abstract**

In structure-based virtual screening, it is often necessary to evaluate the binding free energy of protein–ligand complexes by considering not only molecular conformations but also how these structures shift and rotate in space. The number of possible combinations grows rapidly and can become overwhelming. While classical computing has limitations in this context, quantum computing offers a promising alternative due to its inherent parallelism. In this study, we introduce a quantum machine learning approach that encodes molecular information into quantum states and processes them using parameterized quantum gates. The model is implemented and trained using PyTorch, and its performance is evaluated under three settings: ideal simulation, limited-shot sampling, and simulations with quantum noise. With six quantum circuit units, the model achieves an RMSD of 2.37 kcal/mol and a Pearson correlation of 0.650. Even when using 100,000 shots, the predictions remain consistent, indicating that the model is compatible with near-term quantum hardware. Although noise slightly reduces accuracy, the ranking of ligand affinities remains largely unchanged. These findings point to a practical and scalable strategy that balances robustness and predictive power, offering a viable path to accelerate virtual screening through moderately deep quantum circuits.

**Introduction**

Developing new drugs requires tremendous effort in terms of both cost and time to reach final approval. Structure-based virtual screening (SBVS) aims to accelerate early-stage drug discovery by leveraging the three-dimensional structures of target proteins to identify ligands with strong binding affinity (*1, 2*). Molecular dynamics (MD) simulations with explicit solvent molecules can be used to compute the binding free energy ($\Delta G_{bind}$) of protein-ligand complexes, often achieving good agreement with experimental data (*3*). However, applying MD simulations to large compound libraries is computationally prohibitive. As a result, SBVS generally relies on approximated binding scores instead of directly computing $\Delta G_{bind}$, enabling faster screening of massive chemical databases (*4, 5*).

Scoring functions in SBVS are typically categorized into two groups: physics-based and machine learning–based approaches (*6-12*). Force field-based scoring methods utilize atomic-level interactions such as electrostatic and van der Waals forces. These methods require accurate atomic coordinates of both the ligand and the protein. However, proteins often undergo conformational changes, especially within the binding pocket when a ligand binds. Since the atomic structures of these induced-fit complexes are not always available, the accuracy of force field–based methods is inherently limited.

One way to overcome this limitation is to encode protein–ligand complexes with both atom types and spatial coordinates. Ligands are then evaluated based on how well they complement the protein in space and chemistry. To further reduce sensitivity to small atomic shifts, molecular density functions are employed, treating atomic occupancy in a manner similar to pixel intensity in image recognition (*13, 14*).

Modern compound libraries illustrate the immense scale of chemical space. For example, ZINC20 contains roughly $10^9$ purchasable molecules with available 3D structures (*15*), and GDB-17 includes over $10^{11}$ small molecules with up to 17 atoms (*16*). Estimates suggest that as many as $10^{60}$ organic molecules with molecular weights under 500 Da could be relevant for drug discovery (*17*). Exhaustively evaluating binding scores for such a vast number of molecules is beyond the capability of conventional computing systems due to limitations in processing speed and energy consumption, especially for machine learning models with high computational demands.

Quantum computing offers a promising solution to this scalability bottleneck. A quantum circuit with $n$ qubits can represent a superposition of $2^n$ quantum states simultaneously (*18*). In our approach, these quantum states encode the structural features of protein-ligand binding sites. Quantum gates, which are unitary transformations, evolve the system by altering the amplitudes of these quantum states. A quantum state can be represented in binary form as $(a_{n-1},\ldots, a_m,\ldots, a_0)$, where $a_i$ denotes the state of the $i^{th}$ qubit. When a single-qubit gate is applied to the $m^{th}$ qubit, only the corresponding bit $a_m$ is changed, while the others remain unchanged. As a result, the operation produces a linear combination of all $2^n$ basis states, modifying all amplitudes in parallel. This ability to transform all states simultaneously stands in sharp contrast to classical logic circuits, where each gate operates on a single bit and only one state can be represented at a time (*19*). This parallelism offers a powerful computational advantage, particularly for high-throughput data processing tasks such as SBVS.

Recent advances in quantum hardware make this approach increasingly feasible. Leading platforms include IBM's superconducting qubits, Google's Sycamore processor, China's photonic Jiuzhang system, IonQ's trapped-ion architecture, and Rigetti's superconducting chips (*20-24*). It's still challenging to encode large-scale classical data into quantum states. And quantum noise continues to limit the depth and reliability of circuits. While some researchers focus on improving the hardware itself, others are looking into software methods to reduce errors.

In this study, we assume the availability of ideal quantum hardware and efficient encoding of classical data into quantum states. Under such conditions, if a quantum circuit can estimate the $\Delta G_{bind}$ of a single protein-ligand complex using *n* qubits, then theoretically, adding just 200 more qubits could enable the simultaneous evaluation of up to $10^{60}$ complexes. Prior research has already demonstrated the applicability of quantum machine learning (QML) in drug discovery, including quantum generative adversarial networks for ligand generation (*25*) and quantum support vector machines for molecular classification (*26*). Moreover, quantum convolutional neural networks have shown strong performance in image recognition tasks, further motivating their adaptation to biomolecular data (*27-35*).

Building on these advances, we propose a novel quantum machine learning framework using parameterized quantum circuits to estimate the binding free energies of protein-ligand complexes within SBVS.

**Methods**
**Dataset Preparation for Training and Validation.** This study utilized the PDBbind v2020 dataset, which includes 19,443 protein-ligand complexes with experimentally measured binding affinities reported as $pK_d$ values (*36*). The dataset was partitioned into a training set and a test set. The test set consisted of the Core subset, containing 285 complexes, while the remaining 19,158 complexes were used for training (*37*). To construct the input features for the quantum circuit model, atomic coordinates and atom types of each protein-ligand complex were converted into 512 occupancy values. They represent the three-dimensional structure of each complex and serve as input to the quantum circuit. The $pK_d$ values were converted to binding free energies (in kcal/mol) using the following thermodynamic relation:

$$\Delta G_{bind} = -\ln 10 * RT * pK_d \quad (1)$$

where $R = 1.987 \times 10^{-3}$ kcal·mol$^{-1}$·K$^{-1}$ is the gas constant and $T = 298$ K is the standard temperature. To encode structural features spatially, a cubic grid with a side length of 16 Å was centered on the ligand. Each complex was voxelized into a $32 \times 32 \times 32$ grid aligned along the Cartesian axes. Each voxel represents the occupancy of atoms within that spatial region. To reduce dimensionality, max pooling with a kernel size of $8 \times 8 \times 8$ was applied, resulting in a $4 \times 4 \times 4$ grid. The occupancy value of an atom at a given voxel was computed using:

$$O_{atom}(r) = \begin{cases} e^{-2r^2} & r < 1 \\ \left(\dfrac{3-2r}{e}\right)^2 & 1 \leq r < 1.5 \\ 0 & r \geq 1.5 \end{cases} \quad (2)$$

where *r* is the normalized distance between the voxel center and the atom center, defined as the ratio to the atom's van der Waals radius. The decay function may take a Gaussian or exponential form (*13*). Atoms were categorized into four types, namely carbon (C), nitrogen (N), oxygen (O), and others. The corresponding van der Waals radii were 1.9 Å for carbon, 1.8 Å for nitrogen, 1.7 Å for oxygen, and 2.0 Å for the remaining atom types. Atom types were treated separately for proteins and ligands, resulting in eight distinct input channels. For each complex, 512 voxel-wise occupancy values were calculated. These values were then scaled such that the sum of the squared occupancies was normalized to 0.5 for protein atoms and 0.5 for ligand atoms, respectively, to ensure balanced quantum state encoding.

**Quantum Circuit Architecture.** To implement quantum machine learning for structure-based virtual screening, we adopted a modular design composed of repeated functional blocks, each referred to as a QMLunit, following the framework introduced by Zhao et al. (Zhao, 2021). As shown in Figure 1a, each QMLunit consists of two sequential layers. The first layer, called $L_{par}$, applies RX and RZ rotation gates. The detailed structure of the $L_{par}$ subcircuit is illustrated in Figure 1b.

The second layer, $L_{breaker}$, consists of controlled-NOT (CNOT) gates. Each $L_{breaker}$ configuration includes two layers, each with four CNOT gates. In $L_{breaker}(n)$, the $n^{th}$ qubit is excluded from being a target, resulting in a unique entanglement topology for each index $n$. The full configurations of $L_{breaker}(0)$ through $L_{breaker}(8)$ are shown in Figure 2. These diverse entanglement patterns across QMLunits enhance the circuit's ability to model complex interdependencies within the input features.

By sequentially stacking multiple QMLunits, each of which consists of one $L_{par}$ layer followed by one $L_{breaker}$ layer, the circuit achieves expressive depth. It utilizes a total of nine qubits. Three of them ($T_2$, $T_1$, $T_0$) are designated for encoding atom type information, and the remaining six ($X_1$, $X_0$, $Y_1$, $Y_0$, $Z_1$, $Z_0$) represent the spatial positions of atoms in a discretized three-dimensional voxel grid. This configuration enables the model to incorporate both chemical and geometric features of the protein–ligand complex into its quantum state.

To reduce measurement overhead and minimize the number of shots required during inference, only a single qubit, namely the $0^{th}$ qubit, is measured after circuit execution. The probability distribution of this qubit is used to estimate the binding free energy. This approach enables efficient sampling while preserving prediction accuracy and is particularly suitable for near-term quantum devices with limited readout bandwidth.

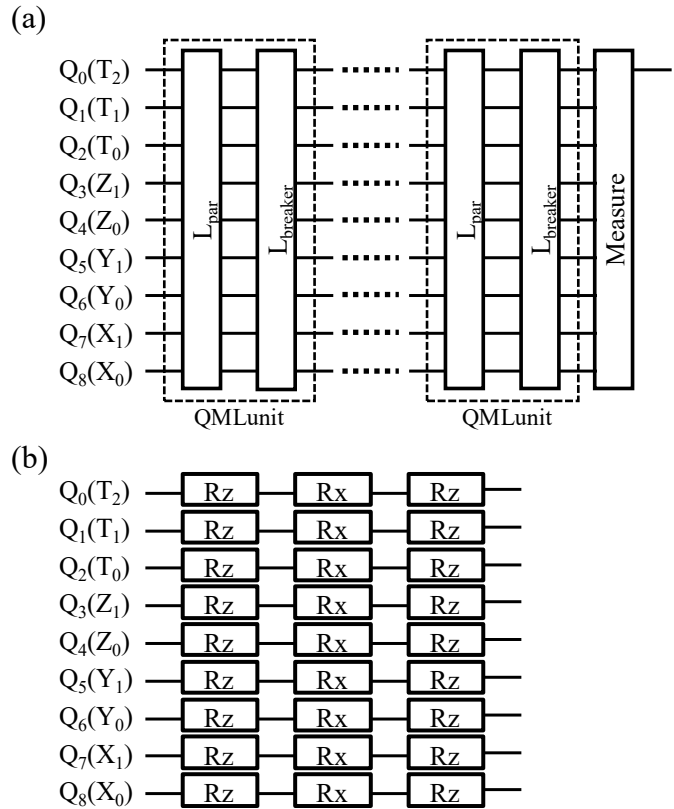

**Figure 1.** (a) Overall quantum circuit architecture, consisting of multiple stacked QMLunits. Each QMLunit includes two consecutive layers: a parameterized single-qubit rotation layer ($L_{par}$) and an entanglement layer ($L_{breaker}$). (b) Internal structure of the $L_{par}$ layer, where each qubit is acted on by a pair of RX and RZ rotation gates, each with its own trainable parameter.

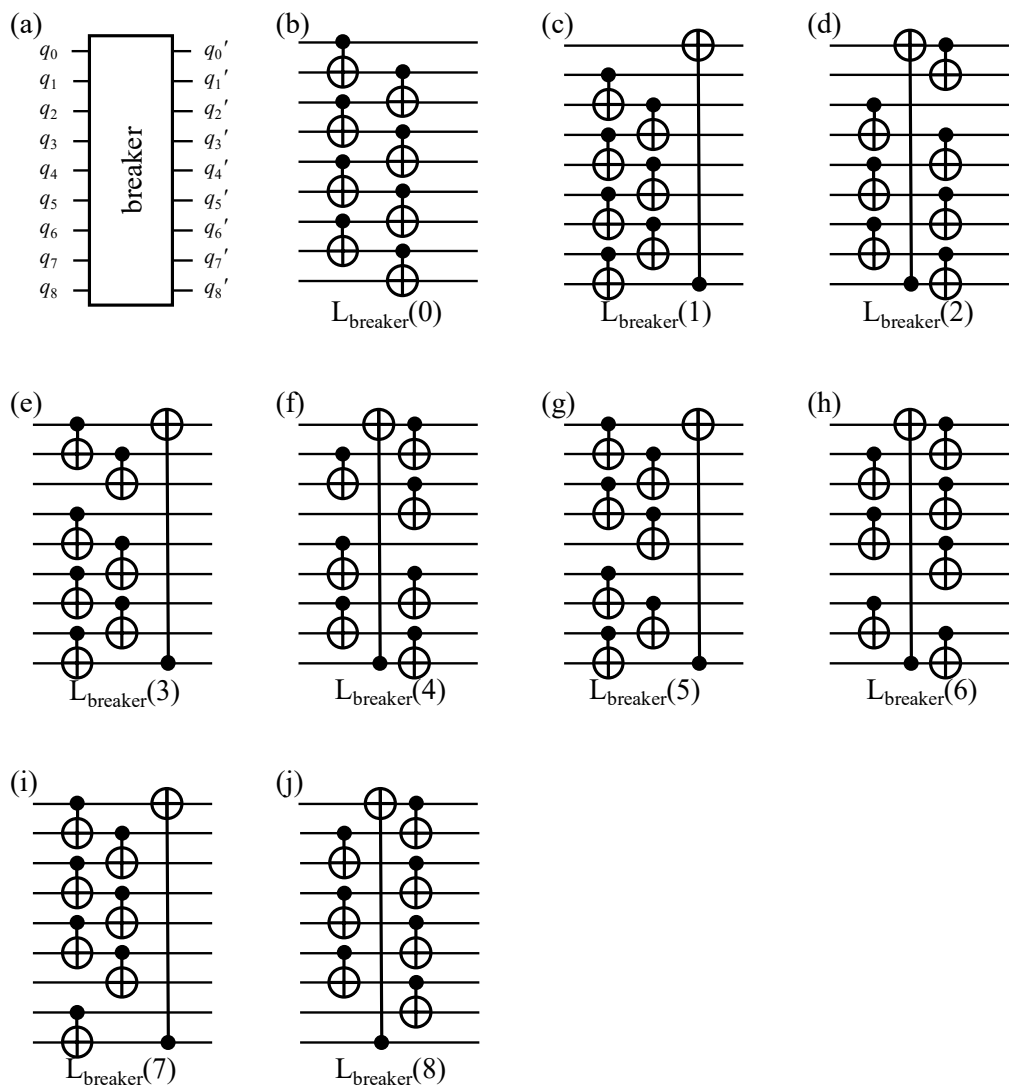

**Figure 2.** (a) Overview of the entanglement module $L_{breaker}$, designed to introduce controlled entanglement across qubits. (b–j) Configurations of $L_{breaker}(0)$ through $L_{breaker}(8)$, where each circuit consists of two layers of CNOT gates. In each $L_{breaker}(n)$, the $n^{th}$ qubit is explicitly excluded from being a target. Filled dots indicate control qubits, and symbols indicate target qubits.

**Training Protocol.** The quantum circuit model was trained using the PDBbind v2020 training set. For each protein-ligand complex, a normalized occupancy vector was encoded into a quantum state and passed through the parameterized quantum circuit. The output of the circuit consisted of two values per sample, corresponding to expectation values measured from the final quantum state. The predicted binding free energy was computed by applying a scaling factor of 100 to the difference between the probabilities of measuring $|0\rangle$ and $|1\rangle$ on qubit 0.

$$\Delta G_{bind} = 100(y_0 - y_1) \tag{3}$$

where $y_0$ and $y_1$ represent the probabilities of obtaining measurement outcomes $|0\rangle$ and $|1\rangle$, respectively, on qubit 0.

The model was optimized to minimize the mean squared error between the predicted $\Delta G_{bind}$ values and the experimental measurements.

We tested four learning rates: $10^{-7}$, $10^{-6}$, $10^{-5}$, and $3\times10^{-5}$. For each setting, three independent training runs were carried out using different random parameter initializations, resulting in a total of 12 trials. The model with the lowest RMSD on the training set was selected for evaluation on the test set. Its parameters were retained for all downstream inference tasks, including full probability simulation, finite-shot sampling, and noisy circuit evaluation.

**Evaluation Protocol.** The trained quantum circuit was evaluated using the PDBbind v2020 core set. We used three evaluation modes, each corresponding to a different level of quantum hardware capability: full-probability simulation, finite-shot sampling, and noisy simulation.

In the full probability simulation, the complete output quantum state was computed by simulating the full quantum state vector in PyTorch. The output amplitudes were used to calculate exact expectation values, which were then transformed into predicted binding free energies using Equation (3).

In the sampling-based evaluation, we simulated the behavior of a practical quantum device by sampling measurement outcomes from the output state. To reduce computational overhead, we measured only a single qubit (the $0^{th}$ qubit), which was then converted into the predicted binding free energy via Equation (3). We tested the model using 10,000 and 100,000 measurement shots to evaluate the impact of sampling noise. For the noisy simulation, we introduced amplitude damping (with a damping rate of 0.001) and depolarizing noise (with a probability of 0.0005).

**Results**

**Estimation of $\Delta G_{bind}$ Using Full Probability.** To establish a performance under ideal quantum conditions, we first evaluated the model using full probability inference. Figures 3 and 4 illustrate the model's prediction accuracy on both training and test sets. On the training set, the RMSD ranged from 2.19 to 2.59 kcal/mol across circuits of different depths. The PCC steadily improved with circuit depth, reaching a maximum of 0.505 when six QMLunits were used.

The test-set performance was with six QMLunits, yielding an RMSD of 2.37 kcal/mol and a PCC of 0.650. Beyond this depth, both metrics began to decline slightly, indicating diminishing returns. These results demonstrate that moderately deep quantum circuits, particularly those with five to six QMLunits, strike a favorable balance between expressiveness and generalization in predicting protein–ligand binding free energies under noise-free, full-statevector simulation conditions.

**Estimation Using Sampling-Based Inference.** To assess the feasibility of deploying the model on real quantum devices, we evaluated its performance under sampling-based inference. Rather than simulating the whole quantum state, we approximated the output by repeatedly measuring a single qubit. Specifically, only the $0^{th}$ qubit was measured. We examined shot counts of 10,000, 100,000, and 1,000,000 to assess the impact of sampling noise on prediction accuracy.

As illustrated in Figure 3, using 100,000 shots resulted in a slight increase in RMSD

compared to full-probability inference. The test-set RMSD ranged from 2.39 to 2.74 kcal/mol, depending on the number of QMLunits. The PCC reached a maximum of 0.632 when six QMLunits were used. In contrast, reducing the shot count to 10,000 led to a noticeable increase in RMSD and a decrease in PCC. Increasing the shot count to 1,000,000 produced results that closely matched those of the full-probability simulation (data not shown).

These findings confirm that while lower shot counts introduce variability in energy estimation, the model retains robustness in maintaining the relative ranking of ligand affinities. This robustness makes the model suitable for practical applications in structure-based screening, even under conditions where quantum measurements are constrained.

**Estimation Under Noisy Simulation.** To further assess the model's robustness under realistic quantum conditions, we evaluated its performance in the presence of simulated quantum noise. Two types of noise channels were applied after each quantum gate: an amplitude damping channel with a damping rate of 0.001 to simulate energy dissipation and a depolarizing channel with a depolarization probability of 0.0005 to model random gate errors. The circuit output was estimated based on the resulting probability distributions derived from the noisy quantum state.

As shown in Figure 3, the introduction of noise led to an increase in RMSD, particularly as the circuit depth increased. When using six QMLunits, the RMSD rose from 2.37 kcal/mol under full-probability inference to approximately 2.97 kcal/mol under noisy simulation. Despite this reduction in absolute accuracy, the PCC remained relatively stable, reaching 0.653, which is comparable to the value in the noise-free case.

These results indicate that the proposed circuit architecture keeps its ability to capture relative binding even under noise conditions. The PCC values suggest that ligand ranking is still a reliable approach. This robustness supports the model's practical use on near-term quantum hardware, where decoherence and gate imperfections are expected to affect circuit performance.

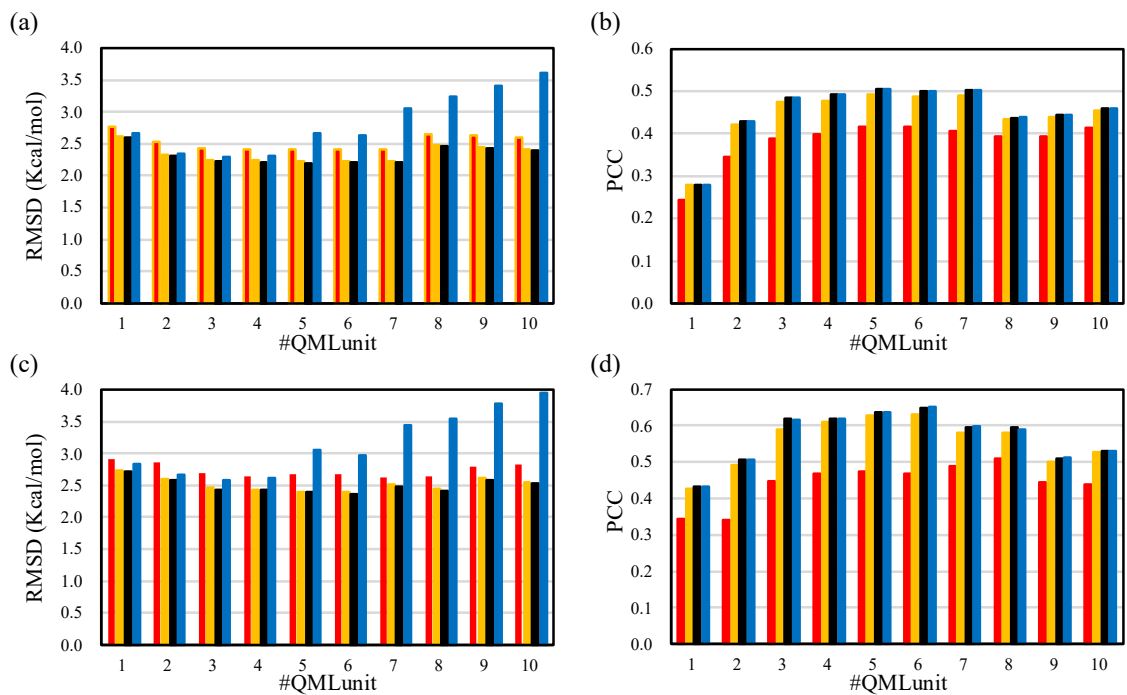

**Figure 3.** Evaluation of model performance across different QMLunit depths under three inference conditions: full-probability simulation (black), sampling-based inference using 10,000 shots (red) and 100,000 shots (orange), and noisy simulation with amplitude damping (rate = 0.001) and depolarizing noise (probability = 0.0005) (blue). Panels (a) and (b) show results on the training set, while panels (c) and (d) show results on the test set. Subfigures (a) and (c) report the RMSD between predicted and experimental $\Delta G_{bind}$ values, while (b) and (d) present the PCC, which reflects the ranking accuracy of ligand binding affinities. The horizontal axis indicates the number of QMLunits.

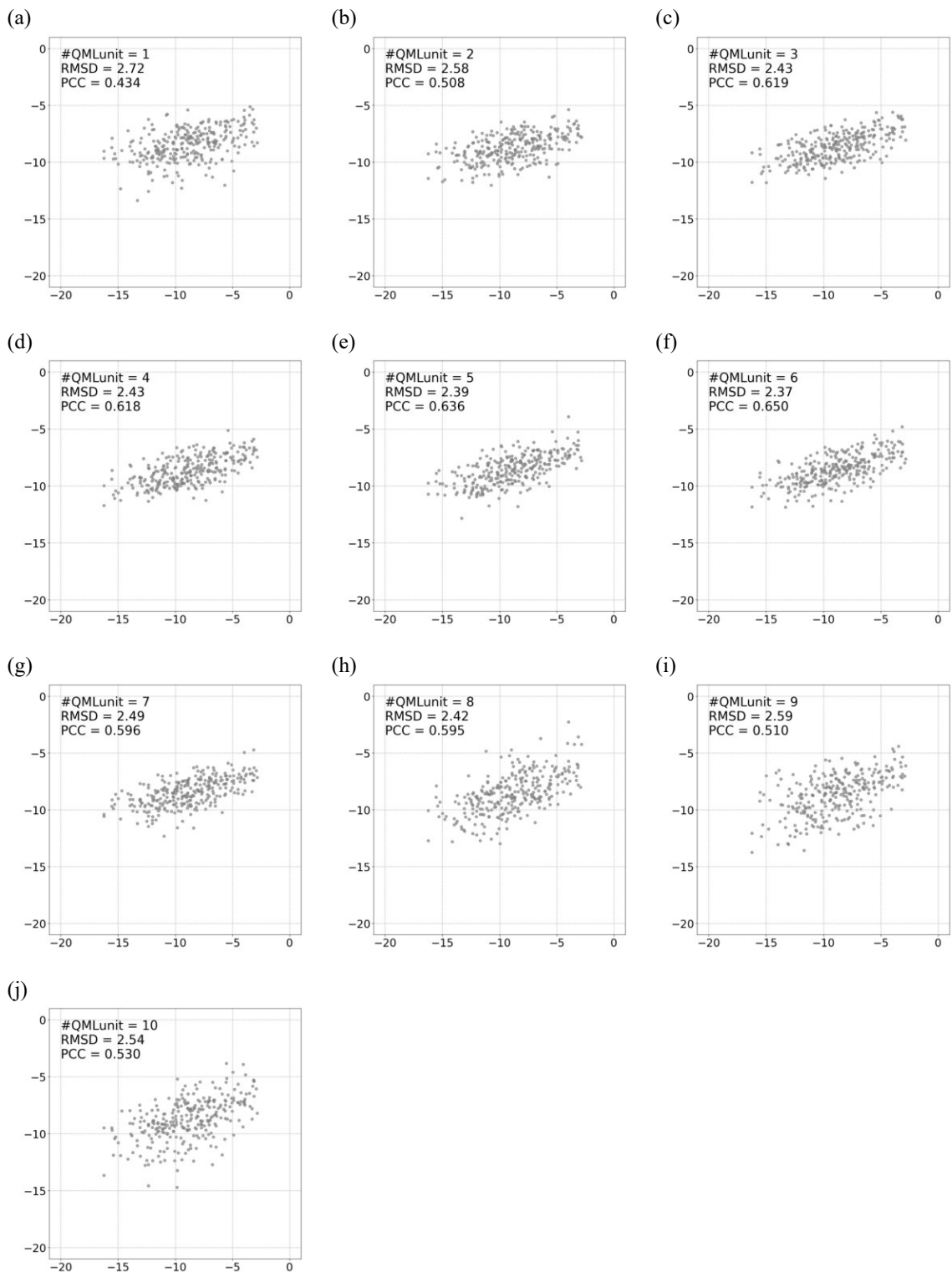

**Figure 4.** Scatter plots of predicted versus experimental ∆$G_{bind}$ on the test set under full-probability inference for circuits with varying numbers of QMLunits. Each panel corresponds to a different QMLunit depth, ranging from 1 to 10. Predicted values are generated by the trained quantum model and compared against experimental values. The diagonal line in each panel represents perfect agreement. As circuit depth increases, the predictions show improved alignment with the experimental data, peaking at six QMLunits, which yields the lowest root mean square deviation (RMSD = 2.37 kcal/mol) and the highest Pearson correlation coefficient (PCC = 0.650).

**Discussions**

    **Advantages of quantum algorithms.** The motivation for adopting quantum algorithms in this study stems from the exponential complexity of the conformational search space in protein-ligand binding. Classical methods struggle with the large number of possible ligand poses, including translations, rotations, and internal flexibility, particularly when receptor conformational changes are also taken into account. Quantum computing offers a distinct perspective. It enables the encoding of high-dimensional molecular states and the simultaneous exploration of vast search spaces, made possible by quantum superposition and entanglement.

    Although current quantum hardware is still in its early stages, parameterized quantum circuits already demonstrate promising capabilities for modeling complex molecular interactions. These circuits can compress intricate structural features into a compact quantum representation. The model appears to maintain a reasonably accurate ligand ranking, even when the number of shots is limited or noise is present. That suggests even shallow quantum circuits can pick up on functional spatial and chemical patterns. As hardware improves, deeper circuits and more complex designs may push accuracy further, potentially even beyond what classical machine learning can achieve in large-scale virtual screening.

    **PyTorch-Based Quantum Circuit Simulation.** To enable end-to-end training and gradient-based optimization of the quantum circuit model, we developed a custom simulation framework implemented in PyTorch. Rather than relying on existing quantum simulators, we represented the quantum state as a tensor and explicitly constructed the full unitary transformation of the circuit by computing Kronecker products of parameterized single-qubit and multi-qubit gate matrices. The final state vector was obtained via matrix multiplication, and the predicted binding free energy was derived from the expectation value of a designated observable.

    This approach allowed integration with PyTorch's automatic differentiation engine, enabling gradients of the loss function to propagate through the entire quantum computation graph. Leveraging tensor algebra and GPU acceleration, the simulation framework supported efficient training across various circuit configurations and batch sizes. This setup provided high flexibility for testing different architectural designs, initialization schemes, and loss functions.

    **Effect of Sampling with Finite Shots**. Sampling-based inference estimates the expectation value of an observable by repeatedly measuring the quantum system with a limited number of shots. This process introduces randomness, which can cause the predicted $\Delta G_{bind}$ values to vary slightly. Using more shots can help reduce this variation and make the results more precise, but the overall improvement in accuracy is usually slight. This highlights the importance of having a good readout strategy.

    In this work, we chose to measure only the $0^{th}$ qubit to keep the sampling cost low, since measuring more qubits would have required many more shots to get the same level

of accuracy. Additionally, increasing the shot count raises computational costs or device usage fees, so it's essential to consider how much improvement in prediction reliability is achieved. Based on our results, using 100,000 shots appears to be a good middle ground. It provides stable and consistent predictions without requiring excessive resources. By balancing shot count and model robustness, the proposed framework achieves reliable performance while maintaining compatibility with near-term quantum devices.

**Effect of Quantum Noise.** Quantum noise presents a fundamental challenge in the practical implementation of near-term quantum algorithms. In this study, we simulated two commonly used noise models: amplitude damping and depolarizing noise. We used these noise models to simulate realistic hardware-level errors. Both types of noise were added after every gate. As the circuit became deeper, the total amount of noise accumulated, resulting in a clear drop in prediction accuracy.

Noise pushed the RMSD higher compared to the noise-free case. For instance, with six QMLunits, the RMSD went from 2.37 kcal/mol under ideal conditions to about 2.97 kcal/mol when noise was added. Although the overall prediction accuracy decreased, the PCC remained relatively stable across different circuit depths. The stability of the PCC indicates that the model effectively preserved the ranking order of ligand affinities, even in the presence of decoherence and gate imperfections.

These results demonstrate that the circuit's design plays a crucial role in mitigating quantum noise. Using shallow QMLunits, keeping entanglement local, and measuring only one qubit all seem to help make the model more robust. This design is particularly suitable for NISQ devices and supports the practical feasibility of applying quantum circuits to real-world molecular property prediction tasks.

**Effect of QMLunit Depth on Predictive Accuracy.** Circuit depth plays a significant role in parameterized quantum models. Here, we varied the number of QMLunits systematically and assessed predictive performance under three inference conditions. We observed that as the circuit depth increased from one to six QMLunits, the model consistently improved in both RMSD and PCC during full-probability simulation.

However, adding more than six QMLunits did not lead to meaningful improvements. This drop in performance may be related to the barren plateau problem. The lack of non-linear activation in quantum circuits could exacerbate the issue even further.

These results suggest that although deeper circuits are better at capturing complex molecular features, adding too much depth can lead to overfitting. Based on our findings, using five to six QMLunits seems to strike a good balance between expressiveness and stability, especially on NISQ hardware.

**Parallel Estimation of Multiple Protein–Ligand Complexes.** In addition to enhancing predictions for individual complexes, the proposed quantum architecture also supports the simultaneous prediction of multiple protein-ligand complexes. This capability leverages the intrinsic parallelism of quantum computing, providing a practical mechanism for accelerating SBVS.

In its standard form, the $QC_{bind}$ circuit processes a single complex using nine qubits. To enable parallel estimation, we introduce $m$ ancillary qubits that serve as indices for multiple inputs. With this extension, the total number of qubits becomes $m + 9$, and the $QC_{bind}$ operation is applied conditionally across $2^m$ distinct input states in parallel, as illustrated in Figure 5.

Let $U_{2^9*2^9}$ denote the unitary matrix corresponding to QC$_{bind}$. When extended to $m + 9$ qubits, the overall transformation becomes a block-diagonal unitary matrix, where each diagonal block equals $U_{2^9*2^9}$. The output quantum states $O_{2^9}^j$ for each of the $2^m$ inputs can thus be computed simultaneously using the matrix relation:

$$\begin{bmatrix} O_{2^9}^0 \\ O_{2^9}^1 \\ \dots \\ O_{2^9}^{2^m-1} \end{bmatrix}_{2^{m+9}} = \begin{bmatrix} U_{2^9*2^9} & 0 & \dots & 0 \\ 0 & U_{2^9*2^9} & \dots & 0 \\ \dots & \dots & \dots & \dots \\ 0 & 0 & \dots & U_{2^9*2^9} \end{bmatrix}_{2^{m+9}*2^{m+9}} \begin{bmatrix} I_{2^9}^0 \\ I_{2^9}^1 \\ \dots \\ I_{2^9}^{2^m-1} \end{bmatrix}_{2^{m+9}} \quad (4)$$

This matrix multiplication can also be reorganized in a batch-friendly form:

$$\begin{bmatrix} O_{2^9}^0 & O_{2^9}^1 & \dots & O_{2^9}^{2^m-1} \end{bmatrix}_{2^9*2^m} = U_{2^9*2^9} \begin{bmatrix} I_{2^9}^0 & I_{2^9}^1 & \dots & I_{2^9}^{2^m-1} \end{bmatrix}_{2^9*2^m} \quad (5)$$

This formulation enables the use of a single QC$_{bind}$ circuit to process multiple inputs simultaneously through vectorized matrix operations, allowing efficient inference of multiple protein–ligand complexes on classical hardware during the prototyping phase. This design provides a scalable foundation for high-throughput virtual screening workflows, underscoring the flexibility and extensibility of the proposed quantum circuit architecture.

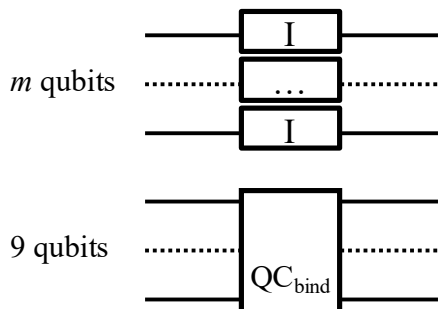

**Figure 5.** Conceptual diagram illustrating the parallel estimation of $\Delta G_{bind}$ values for multiple protein–ligand complexes using a shared quantum circuit (QC$_{bind}$). $m$ ancillary qubits index each complex. Each complex is independently encoded into a quantum state and processed in parallel. The outputs of the circuit are measured separately to obtain $\Delta G_{bind}$ predictions.

**Conclusions**

This study introduces a parameterized quantum circuit model for predicting protein–ligand $\Delta G_{bind}$ in SBVS. The circuit is composed of multiple QMLunits, each combining single-qubit rotations with entanglement layers to balance expressive power and practical feasibility on NISQ devices. The model encodes key molecular features using nine qubits, with three assigned to atom types and six to spatial coordinates. We evaluated the model under full-probability simulation, finite-shot sampling, and noisy execution. Across all settings, it maintained reliable predictive performance. Circuits with five to six QMLunits produced consistent RMSD and PCC scores. Even under noisy conditions, the model preserved the relative ranking of ligand affinities, indicating robustness suitable for realistic hardware. In addition, we explored a parallel estimation method that uses ancillary qubits to process multiple protein–ligand complexes within a single circuit. This approach offers scalability and demonstrates the flexibility and potential of the proposed quantum framework for high-throughput SBVS.

**Competing interests**

The authors declare no competing interests.

**Data and Software Availability**

The data supporting the findings of this study are openly available on GitHub at the following URL: https://github.com/peikunyang/13_QCADD_QDL.